\documentclass[traditabstract,final]{aa}
\pdfoutput=1
\usepackage{amsmath}
\usepackage{natbib}
\usepackage{txfonts}
\usepackage{alltt}

\ifx\pdftexversion\undefined
\usepackage[dvips]{graphicx}
\usepackage[dvips]{hyperref}

\else
\usepackage[pdftex]{graphicx}
\usepackage{epstopdf}
\usepackage[pdftex,colorlinks=false,pdfstartview=FitV,urlcolor=pdfwww,
  bookmarks=true,bookmarksnumbered=true]{hyperref}

\fi
\usepackage{url}
\usepackage{framed}
\usepackage{natbib,ifthen}

\bibpunct{(}{)}{;}{a}{}{,} 

\newcommand{\deleted}[1]{{\bf\ (DELETED TEXT)}}

\newcommand{\lmax}{\ensuremath{l_{\text{max}}}}
\newcommand{\nside}{\ensuremath{N_{\text{side}}}}
\newcommand{\ntheta}{\ensuremath{N_{\vartheta}}}
\newcommand{\alm}{\ensuremath{a_{lm}}}
\newcommand{\libpsht}{\texttt{libpsht}}
\newcommand{\Libpsht}{\texttt{Libpsht}}
\newcommand{\lamlm}{\ensuremath{{{}_s\lambda_{lm}(\vartheta)}}}
\newcommand{\fnurl}[1]{\footnote{\url{#1}}}
\newcommand{\compl}[1]{\ensuremath{\mathcal{O}(#1)}}
\newcommand{\sE}{{}_{|s|}E}
\newcommand{\sB}{{}_{|s|}B}
\newcommand{\sElm}{\sE_{lm}}
\newcommand{\sBlm}{\sB_{lm}}
\newcommand{\half}{{\textstyle \frac{1}{2}}}

\begin{document}

\title {Libpsht -- algorithms for efficient spherical harmonic transforms}
\author {M.~Reinecke \inst{1}}
\institute {Max-Planck-Institut f\"ur Astrophysik, Karl-Schwarzschild-Str.~1, 85741 Garching, Germany}

\offprints {M.~Reinecke,\\ \email{martin@mpa-garching.mpg.de}}

\date{Received 11 October 2010 / Accepted 1 November 2010}

\abstract {\Libpsht\ (or ``library for performant spherical
harmonic transforms'') is a collection of algorithms for efficient
conversion between spatial-domain and spectral-domain representations of data
defined on the sphere. The package supports both transforms of scalars and
spin-1 and spin-2 quantities, and can be used for a wide range of pixelisations
(including HEALPix, GLESP, and ECP).
It will take advantage of hardware features such as multiple processor cores and
floating-point vector operations, if available. Even without this
additional acceleration, the employed algorithms are among the most efficient
(in terms of CPU time, as well as memory consumption)
currently being used in the astronomical community.

The library is written in strictly standard-conforming C90, ensuring portability
to many different hard- and software platforms, and allowing straightforward
integration with codes written in various programming languages like C, C++,
Fortran, Python etc.

\Libpsht\ is distributed under the terms of the GNU General Public
License (GPL) version 2 and can be downloaded from
\url{http://sourceforge.net/projects/libpsht/}.
 }

\maketitle

\section {Introduction}
Spherical harmonic transforms (SHTs) have a wide range of
applications in science and engineering. In the case of CMB science, which
prompted the development of the library presented in this paper, they are an
essential building block for extracting the cosmological power spectrum from
full-sky maps, and are also used for generating synthesised sky maps in the
context of Monte Carlo simulations.
For all recent experiments in this field, the extraction of spherical harmonic
coefficients has to be done up to very high multipole moments (up
to $10^4$), which makes SHTs a fairly costly operation and therefore a natural candidate
for optimisation. It also gives rise to a number of numerical complications that
must be addressed by the transform algorithm.

The purpose of SHTs is the conversion between functions (or maps) on the sphere
and their representation as a set of spherical harmonic coefficients in the spectral
domain. In CMB science, such transforms are most often needed for quantities
of spin 0 (i.e.\ quantities which are invariant with respect to rotation of
the local coordinate system) and spin 2, because the unpolarised and polarised
components of the microwave radiation are fields of these respective types. For
applications concerned with gravitational lensing, SHTs of spins 1 and 3 are
also sometimes required.

The paper is organised as follows.
The following section lists the underlying equations anf the motivations for
developing \libpsht, as well
as the goals of the implementation. A detailed explanation of the
library's inner workings is presented in Sect.\ \ref{algorithm}, and
Sect.\ \ref{api} contains a high-level overview of the interface it provides.
Detailed studies regarding \libpsht's performance, accuracy, and other
quality indicators were performed, and their results presented in
Sect.\ \ref{benchmarks}. Finally, a summary of the achieved (and not completely achieved)
goals is given, together with an outlook on planned future extensions.

\section{Code capabilities}

\subsection {SHT equations}
A continuous spin-$s$ function $f(\vartheta,\varphi)$ with a spectral band
limit of \lmax\ is related to its corresponding set of spin spherical
harmonic coefficients $_sa_{lm}$ by the following equations:
\begin{eqnarray}
f(\vartheta,\varphi)&=&\sum_{m=-\lmax}^{\lmax} \sum_{l=|m|}^{\lmax} {}_sa_{lm}\, {}_sY_{lm}(\vartheta,\varphi)
\label{ana_true}
\\
_sa_{lm} &=& \int_{\vartheta=0}^\pi \int_{\varphi=0}^{2\pi} d\varphi d\vartheta\, \sin \vartheta\, f(\vartheta,\varphi)\, {}_sY_{lm}^*(\vartheta,\varphi)\,\text{.}
\label{syn_true}
\end{eqnarray}

Eqs.\ \eqref{ana_true} and \eqref{syn_true} are known as \textit{backward}
(or \textit{synthesis}) and \textit{forward} (or \textit{analysis}) transforms,
respectively.
For a discretised spherical map consisting of a vector $\vec p$ of
$N_\text{pix}$ pixels at locations ($\vec \vartheta, \vec \varphi$) and
with (potentially weighted) solid angles $\vec w$, they
change to:
\begin{eqnarray}
{\vec p}_n&=&\sum_{m=-\lmax}^{\lmax} \sum_{l=|m|}^{\lmax} {}_sa_{lm}\, {}_sY_{lm}({\vec\vartheta}_n,{\vec \varphi}_n)
\label{eq_syn}
\\
_s{\hat a}_{lm} &=& \sum_{n=1}^{N_\text{pix}}  {\vec p}_n\, {\vec w}_n\, {}_sY^*_{lm}({\vec\vartheta}_n,{\vec \varphi}_n)\,\text{.}
\label{eq_ana}
\end{eqnarray}
Depending on the choice of \lmax, $\vec w$, and the grid geometry, the ${\vec p} \rightarrow {}_sa_{lm}$
transform may only be approximate, which is indicated by choosing the identifier
$\hat a$ instead of $a$.

The main purpose of the presented code is the efficient implementation
(regarding CPU time as well as memory consumption) of Eqs.\ (\ref{eq_syn}) and
(\ref{eq_ana}) for scalars as well as tensor quantities of spins $\pm1$ and $\pm2$.

\subsection{Design considerations}
\label{goals}
At present, several SHT implementations are in use within the CMB community,
like those distributed with the Fortran and C++ implementations of
HEALPix\fnurl{http://healpix.jpl.nasa.gov} \citep{gorski-etal-2005},
GLESP\fnurl{http://glesp.nbi.dk} \citep{doroshkevich-etal-2005},
\texttt{s2hat}\fnurl{http://www.apc.univ-paris7.fr/~radek/s2hat.html},
\texttt{spinsfast}\fnurl{http://www.physics.miami.edu/~huffenbe/research/spinsfast} \citep{huffenberger-wandelt-2010}, \texttt{ccSHT}\fnurl{http://crd.lbl.gov/~cmc/ccSHTlib/doc},
and IGLOO\fnurl{http://www.cita.utoronto.ca/~crittend/pixel.html}
\citep{crittenden-turok-1998}; they offer varying
degrees of performance, flexibility and feature completeness. The main design
goal for \libpsht\ was to provide a code which performs better
than the SHTs in these packages and is versatile enough to serve as a drop-in
replacement for them.

This immediately leads to several technical and pragmatic requirements which the
code must fulfill:
\begin{description}
\item[\textit{Efficiency:}] The transforms provided by the library must perform at
least as well (in terms of CPU time and memory) as those in the available packages,
on the same hardware. If possible, all available hardware features (like multiple
CPU cores, vector arithmetic instructions) should be used.
\item[\textit{Flexibility:}] The library needs to support all of the above discretisations of the sphere;
     in addition, it should allow transforms of partial maps.
\item[\textit{Portability:}] It must run on all platforms supported by the
  software packages above.
\item[\textit{Universality:}] It must provide a clear interface that can be conveniently called from all the
      programming languages used for the packages above.
\item[\textit{Completeness:}] If feasible, it should not depend on external libraries, because
      these complicate installation and handling by users.
\item[\textit{Compactness:}] In order to simplify code development, maintenance and
      deployment, the library source code should be kept as small as possible without
      sacrificing readability.
\end{description}
It should be noted that the code was developed as a library for use within
other scientific applications, not as a scientific application by itself. As such
it strives to provide a comprehensive interface for programmers, rather than
ready-made facilities for users.

\subsection{Supported discretisations}

For completely general discretisations of the sphere, the operation count
for SHTs is \compl{\lmax^2N_\text{pix}}\ in both directions, which is not
really affordable for today's resolution requirements.
Fortunately, approaches of lower complexity exist for grids with the
following geometrical constraints:
\begin{itemize}
\item the map pixels are arranged on $N_\vartheta\approx\sqrt{N_\text{pix}}$ iso-latitude rings, where
  the colatitude of each ring is denoted as ${\vec\vartheta}_y$.
\item within one ring, pixels are equidistant in $\varphi$ and have identical
  weight ${\vec w}_y$; the number of pixels in each ring can vary and is
  denoted as $N_{\varphi,y}$, and the azimuth of the first pixel in the ring
  is $\varphi_{0,y}$.
\end {itemize}
Under these assumptions, Eqs.\ (\ref{eq_syn}) and (\ref{eq_ana}) can be written as
\begin{eqnarray}
{\vec p}_{xy}&=&\sum_{m=-\lmax}^{\lmax} \sum_{l=|m|}^{\lmax} {}_sa_{lm}\, {}_s\lambda_{lm}({\vec\vartheta}_y)
 \exp{\left(im\varphi_{0,y} + \frac{2\pi imx}{N_{\varphi,y}}\right)}
\label{syn_used}
\\
{}_s{\hat a}_{lm} &=& \sum_{y=0}^{N_\vartheta-1} \sum_{x=0}^{N_{\varphi,y}-1} {\vec p}_{xy}\, {\vec w}_y\, {}_s\lambda_{lm}({\vec\vartheta}_y)\exp{\left(-im\varphi_{0,y} - \frac{2\pi i m x}{N_{\varphi,y}}\right)} \text{,}
\label{ana_used}
\end{eqnarray}
where $\lamlm:={}_sY_{lm}(\vartheta,0)$.
By subdividing  both transforms into two stages
\begin{eqnarray}
{\vec p}_{xy}&=&\sum_{m=-\lmax}^{\lmax}F_{m,y}\exp{\left(im\varphi_{0,y} + \frac{2\pi imx}{N_{\varphi,y}}\right)}
\quad\text{with}
\label{syn1}
\\
F_{m,y}&:=&\sum_{l=|m|}^{\lmax} {}_sa_{lm}\, {}_s\lambda_{lm}({\vec\vartheta}_y)\text{,\quad and}
\label{syn2}
\\
{}_s{\hat a}_{lm} &=& \sum_{y=0}^{N_\vartheta-1}G_{m,y}\,{}_s\lambda_{lm}({\vec\vartheta}_y)\quad\text{with}
\label{ana1}
\\
G_{m,y} &:=& {\vec w}_y\sum_{x=0}^{N_{\varphi,y}-1} {\vec p}_{xy}\, \exp{\left(-im\varphi_{0,y}-\frac{2\pi i m x}{N_{\varphi,y}}\right)}\text{,}
\label{ana2}
\end{eqnarray}
it becomes obvious that the steps $_sa_{lm}\rightarrow F_{m,y}$ and
$G_{m,y}\rightarrow {}_s{\hat a}_{lm}$ require \compl{N_\vartheta \lmax^2} operations,
while a set of fast Fourier transforms -- with the subdominant complexity of
\compl{N_\vartheta\lmax\log\lmax}~-- can be used for the steps
$F_{m,y}\rightarrow{\vec p}_{xy}$ and ${\vec p}_{xy}\rightarrow G_{m,y}$,
respectively.
For any ``useful'' arrangement of the rings on the sphere, $N_\vartheta$ will be
roughly proportional to \lmax, resulting in \compl{\lmax^3} operations per SHT.

Practically all spherical grids which are currently used in the CMB field
(HEALPix, ECP, GLESP, IGLOO) fall into this category, so \libpsht\ can
take advantage of this simplification. A notable exception is the
``quadrilateral spherical cube'' grid used by the COBE mission, but this is
not really an issue, since COBE's resolution is so low that nowadays even
brute-force SHTs produce results in acceptable time.

For the special case of the ECP grid, which is equidistant in both $\vartheta$ and
$\varphi$, algorithms of the even lower complexity \compl{\lmax^2\log^2\lmax}
have been presented (see, e.g., \citealt{driscoll-healy-1994,wiaux-etal-2007}).
These were not considered for \libpsht, because this would conflict with
the flexibility goal. In addition, it has been observed that the constant
pre\-fac\-tor for these
methods is fairly high, so that they do not offer any performance advantage
for small and intermediate \lmax; for transforms with high \lmax, on the other
hand, their memory consumption is prohibitively high.

\subsection{Transformable quantities}
\label{constraints}
Without limiting the general applicability of the provided SHTs, \libpsht\ imposes
a few constraints on the format of the data it processes. For scalar transforms,
it will only work with real-valued maps (the ubiquitous scenario in CMB physics);
this is not really a limitation, since transforming
a complex-valued map can be achieved by simply transforming its real and imaginary
parts independently and computing a linear combination of the results.

For SHTs with $s\ne 0$, \libpsht\ works on the so-called \textit{gradient} (E)
and \textit{curl} (B) components of the tensor field
\begin{eqnarray}
\sElm &:=& (-1)^H \half ({}_{|s|} a_{lm} + (-1)^s {}_{-|s|} a_{lm}) \\
 i\,\sBlm &:=&  (-1)^H \half ({}_{|s|} a_{lm} - (-1)^s {}_{-|s|} a_{lm})
\end{eqnarray}
where $\sElm^* = (-1)^m \sE_{l,-m}$, $\sBlm^* = (-1)^m \sB_{l,-m}$
and $(-1)^H$ is a sign convention \citep{lewis-2005}. The default setting for
$H$ in \libpsht\ is 1, following the HEALPix convention, but this can be changed
easily if necessary.
On the real-space side, the tensor field is represented as two separate maps,
containing the real and imaginary part, respectively.

In the following discussion, no fundamental distinction exists between
$\sElm$ and $\sBlm$, so for the sake of brevity both will be identified as
${}_s\alm$.

This choice of representation was made because it is compatible with the data
formats used by most existing packages, which simplifies interfacing and data
exchange. It also has the important advantage that any set of spherical harmonic
coefficients is completely specified by its elements with $m>=0$, since
${}_s\alm$ and ${}_sa_{l,-m}$ are coupled by symmetry relations.

\section{Algorithm description}
\label{algorithm}
\subsection{Calculation of the spherical harmonic coefficients}
\label{ylmcalc}
The central building block for every SHT implementation is the quick and
accurate computation of spherical harmonics ${}_sY_{lm}(\vartheta,\varphi)$;
for \libpsht, these need only be evaluated at $\varphi=0$ and will be
abbreviated as \lamlm.

\subsubsection{Scalar case}
For $s=0$, a proven method of determining the coefficients is using the
recursion relation
\begin{eqnarray}
\lambda_{lm}=\cos \vartheta A_{lm} \lambda_{l-1,m}-B_{lm}\lambda_{l-2,m}
\label{eq_ylm}
\end{eqnarray}
with
\begin{eqnarray}
A_{lm} := \sqrt{\frac{4l^2-1}{l^2-m^2}} \quad \text{and} \quad
B_{lm} := \frac{A_{lm}}{A_{l-1,m}}
\label{ylm_precomp}
\end{eqnarray}
in the direction of increasing $l$; the starting
values $\lambda_{mm}$ can be obtained from $\lambda_{00}={(4\pi)}^{-1/2}$ and
the additional recursion relation
\begin{eqnarray}
\lambda_{mm}=-sin\vartheta\sqrt{\frac{2m+1}{2m}}\lambda_{m-1,m-1}
\end{eqnarray}
\citep{varshalovich-etal-1988}. Since two values are required to start the
recursion in $l$, one also uses $\lambda_{m+1,m}=\lambda_{mm}\sqrt{2m+3}$.

Both recursions are numerically stable;
nevertheless, their evaluation is not trivial, since the starting values
$\lambda_{mm}$ can become extremely small for increasing $m$
and decreasing $\sin\vartheta$, such that they can no longer be represented
by the IEEE754 number format used by virtually all of today's computers.
This problem was addressed using the approach described by
\cite{prezeau-reinecke-2010}, i.e.\ by augmenting the IEEE representation with
an additional integer scale factor, which serves as an ``enhanced exponent'' and
dramatically increases the dynamic range of representable numbers.

As is to be expected, this extended range comes at a somewhat higher computational
cost; therefore the algorithm switches to standard IEEE numbers as soon as the
$\lambda_{lm}$ have reached a region which can be safely represented without
a scale factor. From that point on, the recursion in $l$ only requires three
multiplications and one subtraction per $\lambda$ value, which are very cheap
instructions on current microprocessors.

From the above it follows that, depending on the exact parameters of the SHT,
a non-negligible fraction of the \lamlm\ is extremely small, and that all terms
in Eqs.\ (\ref{syn2}) and (\ref{ana1}) containing such values
as a factor do not contribute in a measurable way to the SHT's result, due
to the limited dynamic range of the floating-point format in which the
result is produced. For that reason, \libpsht\ records the index $l_{\text{th}}$,
at which the $|\lamlm|$ first exceed a threshold $\varepsilon_{\text{th}}$ (which
is set to the conservative value of $10^{-30}$) during every $l$ recursion.
The remaining operations will then be carried out for $l\ge l_{\text{th}}$ only.

Another acceleration is achieved by skipping the $l$-recursion for an
$(m,\vartheta)$-tuple entirely, if another recursion was performed before with
$m_c\le m$ and $|\cos\vartheta_c| \le |\cos\vartheta|<1$ that never crossed
$\varepsilon_{\text{th}}$.

In the case that a ring at colatitude $\vartheta$ has a counterpart at
$\pi-\vartheta$ (which is almost always the case for the popular grids),
the \lamlm\ recursion need only be performed for one
of those; the values for the other ring are quickly obtained by the symmetry relation
\begin{eqnarray}
_s\lambda_{lm}(\pi-\vartheta) = (-1)^{l-s} {}_{-s}\lambda_{lm}(\vartheta)\text{.}
\end{eqnarray}
\Libpsht\ will identify such ring pairs automatically and treat them in an
optimised fashion.

\subsubsection{Tensor case}
\label{tensorSHT}
Following \cite{lewis-2005}, \libpsht\ does not literally compute the tensor
harmonics ${}_{\pm s}\lambda_{lm}$, but rather their linear
combinations
\begin{eqnarray}
{}_sF_{lm}^{\pm}:=\half({}_{+s}\lambda_{lm} \pm (-1)^s {}_{-s}\lambda_{lm})\text{.}
\end{eqnarray}
These are not directly obtained using a recursion, but rather by first computing
the spin-0 $\lambda_{lm}$ and subsequently applying spin raising and lowering
operators; see appendix A of \cite{lewis-2005} for the relevant formulae.
Whenever scalar and tensor SHTs are performed simultaneously, this has the
advantage that the scalar $\lambda_{lm}$ can be re-used for both; in CMB science
this scenario is very common, since any transform involving a polarised sky map
requires SHTs for both $s=0$ and $s=\pm2$.

The approach can be extended to higher spins in principle, but for $|s|>2$ the
expressions become quite involved and are also slow to compute. For this reason,
\libpsht's transforms are currently limited to $s=0,\pm1,\pm2$. Should a need
for SHTs at higher $s$ become evident, a method making use of the relationships between
${}_s\lambda_{lm}$ and the Wigner $d^{l}_{mm'}$ matrix elements seems most promising,
since they obey very similar recursion relations (see, e.g., \citealt{kostelec-rockmore-2008}), for which a recent implementation
is available \citep{prezeau-reinecke-2010}.

\subsubsection{Re-using the computed \lamlm}
\label{precompute}
Many packages allow the user to supply an array containing precomputed
\lamlm\ coefficients to the SHT routines, which can speed up the computation
by a factor of up to $\approx 2$.
\Libpsht\ does not provide such an option; this may seem surprising at first,
but it was left out deliberately for the following reasons:
\begin{itemize}
\item Even if supplied externally, the coefficients have to
  be obtained by some method. Reading them from mass storage is definitely not a
  good choice, since they can be computed on-the-fly much more quickly on
  current hardware. Externally provided values can therefore only provide a
  benefit if they are used in at least two consecutive SHTs.
\item When using multiple threads in combination with precomputed \lamlm,
  computation time soon becomes dominated
  by memory access operations to the table of coefficients, because
  the memory bandwidth of a computer does not scale with the number of
  processor cores. This will severely limit the scalability of the code.
  It is safe to assume that this problem will become even more pronounced in the
  future, since CPU power tends to grow more rapidly than memory bandwidth
  \citep{drepper-2007}.
\item For problem sizes with $\lmax\gtrapprox 2000$, when additional performance
  would be most desirable, the precomputed tables
  (whose size is proportional to $N_\vartheta \lmax^2$) become so large that
  they no longer fit into main memory. A conservative estimate for the size
  of the \lamlm\ array at $\lmax=2000$ and $s=0$ is about 8GB; the Fortran HEALPix
  library currently requires 32GB in this situation.
\item \Libpsht\ supports multiple simultaneous SHTs, and will
  compute the \lamlm\ only once during such an operation, re-using them for all
  individual transforms. Since this is done in a piece-by-piece fashion
  (see Sect.\ \ref{loopstructure}), only small
  subsets of size \compl{\lmax}\ need to be stored, which is more or less negligible;
  this approach also implies that the internally precomputed \lamlm\ subsets will most likely
  remain in the CPU cache and therefore not be subject to the memory bandwidth limit.
\end{itemize}

\subsection {Fast Fourier transform}
\Libpsht\ uses an adapted version of the well-known \texttt{fftpack} library
\citep{swarztrauber-1982}, ported to the C language (with a few minor
adjustments) by the author.
This package performs well for FFTs of lengths $N$ whose prime decomposition
only contains the factors 2, 3, and 5. For prime $N$, its complexity degrades
to \compl{N^2}, which should be avoided, so for $N$ containing large prime
factors, Bluestein's algorithm \citep{bluestein-1968} is employed, which allows
replacing an FFT of length $N$ by a convolution (conveniently implemented as a
multiplication in Fourier space) of any length
$\ge2N-1$. By appropriate choice of the new length, the desired \compl{N\log N}
complexity can be reestablished, at the cost of a higher prefactor.

Obviously it would have been possible to use the very popular and efficient
\texttt{FFTW}\fnurl{http://fftw.org} library \citep{frigo-johnson-2005}
for the Fourier transforms, but since this package is fairly heavyweight,
this would conflict with \libpsht's completeness or compactness goals.
Even though \texttt{libfftpack} with the Bluestein modification will be
somewhat slower than \texttt{FFTW}, the efficiency goal is not really
compromised: \libpsht's algorithms have
an overall complexity of \compl{N_\vartheta \lmax^2}, so that the FFT part with
its total complexity of \compl{N_\vartheta\lmax\log\lmax} will (for nontrivial
problem sizes) never dominate the total CPU time.

\subsection{Loop structure of the SHT}
\label{loopstructure}

As Eqs.\ (\ref{syn_used}) and (\ref{ana_used}) and the complexity
of \compl{N_\vartheta \lmax^2} already suggest, a code implementing SHTs
on an iso-latitude grid will contain at least three nested loops which iterate
on $y$, $m$, and $l$, respectively. The overall speed
and/or memory consumption of the final implementation depend sensitively on
the order in which these loops are nested, since different hierarchies are not
equally well suited for precomputation of common quantities used in the
innermost loop.

The choice made for the computation of the \lamlm, i.e.\ to obtain them by
$l$-recursion (see Sect.\ \ref{ylmcalc}) strongly favors placing the
loop over $l$ innermost; so only the order of the $y$ and $m$ loops remains
to be determined.

From the point of view of speed optimisation, arguably the best approach would
be to iterate over $m$ in the outermost loop, then over the ring number $y$, and
finally over $l$. This allows precomputing the $A_{lm}$ and $B_{lm}$ quantities
(see Sect.\ \ref{ylmcalc}) exactly once and in small chunks, for an additional
memory cost of only \compl{\lmax}. However, this method requires additional
\compl{N_{\vartheta} \lmax}\ storage for the intermediate products $F_{m,y}$
or $G_{m,y}$, which is comparable to the size of the input and output data.
This was considered too wasteful, since it prohibits performing SHTs whose
input and output data already occupy most of the available computer memory.

Unfortunately, switching the order of the two outermost loops does not improve
the situation: now, one is presented with the choice of either precomputing the
full two-dimensional $A_{lm}$ and $B_{lm}$ arrays, which is not acceptable for
the same reason that prohibits storing the entire $F_{m,y}$ and $G_{m,y}$
arrays, or recomputing their values over and over, which consumes too much CPU time.

To solve this dilemma, it was decided not to perform the SHTs for a whole map
in one go, but rather to subdivide them into a number of partial transforms,
each of which only deals with a subset of rings (or ring pairs). An appropriate choice for the
number of parts will result in near-optimal performance combined with only small
memory overhead; the most symmetrical choice of $N^{1/2}_{\vartheta}$ leads to
a time overhead of \compl{N^{1/2}_{\vartheta}\lmax^2} and additional memory
consumption of \compl{N^{1/2}_{\vartheta}\lmax}, both of which are small
compared to the total CPU and storage requirements.
\Libpsht\ uses sub-maps of roughly 100 or $N_\vartheta/10$ rings (or ring pairs),
whichever is larger.

The arrangement described above is sufficient for a single SHT. Since
\libpsht\ supports multiple simultaneous SHTs, another loop hierarchy had to be introduced.
A pseudo-code listing illustrating the complete design of the SHT algorithm
(including the subdominant parts performing the FFTs)
is presented in Fig.~\ref{loopstruct}.

\begin{figure}
\begin{framed}
\begin{alltt}
for b = <all submaps or "blocks">
  for y = <all rings in submap b>       // OpenMP
    for j = <all map->a_lm jobs>
      compute G(m,theta_y,j) using FFT  // eq. \eqref{ana2}
    end j
  end y
  for m = [0;mmax]                      // OpenMP
    for l = [m;lmax]
      compute A(l,m) and B(l,m)         // eq. \eqref{ylm_precomp}
    end l
    for y = <all rings in submap b>     // SSE2
      for s = <all required spins>
        for l = [m;lmax]
          compute s_Ylm(theta_y)        // eq. \eqref{eq_ylm}
        end l
      end s
      for j = <all jobs>
        if (a_lm->map job)
          for l = [m;lmax]
            accumulate F(m,theta_y,j)   // eq. \eqref{syn2}
          end l
        else /* map->a_lm job */
          for l = [m;lmax]
            compute a_lm(j)             // eq. \eqref{ana1}
          end l
        endif
      end j
    end y
  end m
  for y = <all rings in submap b>       // OpenMP
    for j = <all a_lm->map jobs>
      compute map(x,y,j) using FFT      // eq. \eqref{syn1}
    end j
  end y
end b
\end{alltt}
\end{framed}
\caption{Schematic loop structure of \libpsht's transform algorithm}
\label{loopstruct}
\end{figure}

\subsection{Further acceleration techniques}

To make use of multiple CPU cores, if available, the algorithm's loop over $m$,
as well as the code blocks performing the FFTs, have been parallelised
using OpenMP\fnurl{http://openmp.org} directives, requesting dynamic
scheduling for every iteration; this is indicated by the \texttt{"OpenMP"} tag
in Fig.~\ref{loopstruct}. As long as the SHTs are large enough, this
results in very good scaling with the number of cores available (see also
Sect.\ \ref{openmp_scaling}).

If the target machine supports the SSE2
instruction set\fnurl{http://en.wikipedia.org/wiki/SSE2} (which is the case for
all AMD/Intel CPUs introduced since around 2003), its ability to perform
two arithmetic operations of the same kind with a single machine instruction
is used to process two ring (pairs) simultaneously, greatly improving the
overall execution speed. The relevant loop is marked with \texttt{"SSE2"}
in Fig.~\ref{loopstruct}. The necessary changes are straightforward for the most
part; however, special attention must be paid to the $l$ recursion, because the
two sequences of \lamlm\ will cross the threshold to IEEE-representable numbers
at different $l_{\text{th}}$. Fortunately this complication can be addressed without
a significant slowdown of the algorithm.

The data types used and the functions called in order to use the SSE2 instruction
set are not part of the C90 language standard, and will therefore not be known
to all compilers and on all hardware platforms. However there is a portable way to
detect compiler and hardware support for SSE2; if the result of this check
is negative, only the standard-compliant floating-point implementation of
\libpsht\ will be compiled, without the necessity of user intervention.

Similarly, the OpenMP functionality is accessed using so-called
\texttt{\#pragma} compiler directives, which are simply ignored by compilers
lacking the required capability, reducing the source code to a single-threaded
version.

Both extensions are supported by the widely used
\texttt{gcc}\fnurl{http://gcc.gnu.org} and Intel compilers.

\section{Programming interface}
\label{api}
Only a high-level overview of the provided functionality can be given here.
For the level of detail necessary to actually interface with the library, the
reader is referred to the technical documentation provided alongside the source code.

\subsection{Description of input/output data}
Due to \libpsht's universality goal, it must be able to accept input data
and produce output data in a wide variety of storage schemes, and therefore
only makes few assumptions about how maps and \alm\ are arranged in memory.

A tesselation of the sphere and the relative location of its pixels in memory is
described to \libpsht\ by the following set of data:
\begin{itemize}
\item the total number of rings in the map; and
\item for every ring
\begin{itemize}
\item the number of pixels $N_{\varphi,y}$;
\item the index of the first pixel of the ring in the map array;
\item the stride between indices of consecutive pixels in this ring;
\item the azimuth of the first pixel $\varphi_{0,y}$;
\item the colatitude ${\vec\vartheta}_y$;
\item the weight factor ${\vec w}_y$. This
is only used for forward SHTs and is typically the solid
angle of a pixel in this ring.
\end {itemize}
\end{itemize}

For \alm, less information is necessary:
\begin{itemize}
\item the maximum $l$ and $m$ quantum numbers available in the set
\item the index stride between $a_{lm}$ and $a_{l+1,m}$;
\item an integer array containing the (hypothetical) index of the element $a_{0,m}$.
\end{itemize}
Due to the convention described in Sect.\ \ref{constraints}, only coefficients
with $m\ge0$ will be accessed.

These two descriptions are flexible enough to describe all spherical grids of
interest, as well as all data storage schemes that are in widespread use
within the CMB community. In combination with a data pointer to the first
element of a map or set of $a_{lm}$, they describe all characteristics of
a set of input or output data that \libpsht\ needs to be aware of.

It should be noted that the map description trivially
supports transforms on grids that only cover parts of the $\vartheta$ range;
rings that should not be included in map synthesis or analysis can simply be
omitted in the arrays describing the grid geometry, speeding up the SHTs on this
particular grid.
More general masks, which affect parts of rings, have to be applied by explicitly
setting the respective map pixels to zero before map analysis or after
map synthesis.

\Libpsht\ will only work on data that is already laid out in main memory according
to the structures mentioned above. There is no functionality to access data
on mass storage directly; this task must be undertaken by the programs calling
\libpsht\ functionality. Given the bewildering variety of formats which are
used for storing spherical maps and \alm\ on disk -- and with the prospect of
these formats changing more or less rapidly in the future -- I/O was not
considered a core functionality of the library.

\subsection{Supported floating-point formats}
In addition to the flexibility in memory ordering, \libpsht\ can also work
on data in single- or double-precision format, with the requirement that
all inputs and outputs for any set of simultaneous transforms must be of the
same type. Internally, however, all computations are performed using
double-precision numbers, so that no performance gain can be achieved by
using single-precision input and output data.

\subsection{Multiple simultaneous SHTs}
\label{joblist}
As already mentioned, it is possible to compute several SHTs at once, independent
of direction and spin, as long as their map and \alm\ storage schemes described
above are identical. This is achieved by repeatedly adding SHT requests, complete
with pointers to their input and output arrays, to a ``job list'', and finally
calling the main SHT function, providing the job list, map information and
\alm\ information as arguments.

\subsection{Calling the library from other languages}
Due to the choice of C as \libpsht's implementation language, the library
can be interfaced without any complications from other codes written in C
or C++. Using the library from languages like Python and Java should also
be straightforward, but requires some glue code.

For Fortran the situation is unfortunately a bit more complicated, because
a well-defined interface between Fortran and C has only been introduced
with the Fortran2003 standard, which is not yet universally supported by
compilers. Due to the philosophy of incrementally building a job list and
finally executing it (see Sect.\ \ref{joblist}), simply wrapping the C functions
using \texttt{cfortran.h}\fnurl{http://www-zeus.desy.de/~burow/cfortran/} glue
code does not work reliably, so that building a portable interface is a nontrivial
task. Nevertheless, the Fortran90 code \texttt{beam2alm} of the \textit{Planck}
mission's simulation package has been successfully interfaced with \libpsht.

\section{Benchmarks}
\label{benchmarks}
In order to evaluate the reliability and efficiency of the library, a series
of tests and comparisons with existing implementations were performed. All of
these experiments were executed on a computer equipped with 64GB of main
memory and Xeon E7450 processors (24 CPU cores overall) running at 2.4 GHz.
\texttt{gcc} version 4.4.2 was used for compilation.

\subsection{Correctness and accuracy}

The validation of \libpsht's algorithms was performed in two stages: first,
the result of a backward transform was compared to that produced with another SHT
library using the same input, and afterwards it was verified that a pair
of backward and forward SHTs reproduces the original \alm\ with sufficient
precision.

All calculations in this section start out with a set of ${}_sa_{lm}$ that
consists entirely of uniformly distributed random numbers in the range $(-1;1)$,
except for the imaginary parts of the ${}_sa_{l,0}$, which must be zero for
symmetry reasons, and coefficients with $l<|s|$, which have no meaning in a
spin-$s$ transform and are set to zero as well.

\subsubsection{Validation against other implementations}
For this test, a set of \alm\ with $\lmax=2048$ was generated using the above
prescription and converted to a $\nside=1024$ HEALPix map by both \libpsht\ and the
Fortran HEALPix facility \texttt{synfast}. This was done for spins 0 and 2,
mimicking the synthesis of a polarised sky map. Comparison of the resulting
maps showed a maximum discrepancy between individual pixels of $1.55\times 10^{-7}$,
and the RMS error was $6.3\times 10^{-13}$, which indicates a very good agreement.

\subsubsection {Accuracy of recovered ${\hat a}_{lm}$}
An individual accuracy test consists of a map synthesis, followed by an analysis
of the obtained map, producing the result ${}_s\hat a_{lm}$.

Two quantities are used to assess the agreement between these two sets of spherical
harmonic coefficients; they are defined as
\begin{eqnarray}
\varepsilon_{\text{rms}} &:=& \sqrt{\frac{\sum_{lm}|{}_sa_{lm}-{}_s\hat a_{lm}|^2}{\sum_{lm}|{}_sa_{lm}|^2}}\quad\text{and}
\\
\varepsilon_{\text{max}} &:=& \max_{lm} \left(|\Re ({}_sa_{lm}-{}_s\hat a_{lm})|,|\Im ({}_sa_{lm}-{}_s\hat a_{lm})|\right)\text{.}
\end{eqnarray}

\begin{figure}
\begin{center}
\includegraphics[width=0.9\columnwidth]{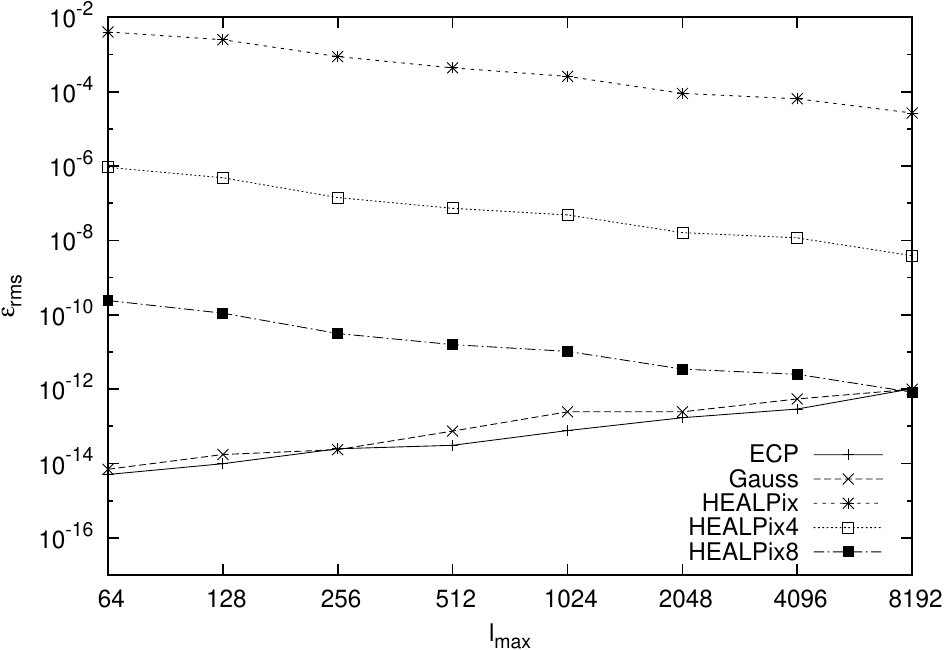}
\end{center}
\caption{$\varepsilon_{\text{rms}}$ for \libpsht\ transform pairs on ECP, Gaussian
and HEALPix grids for various \lmax. Every data point is the maximum error value
encountered in transform pairs of all supported spins. ``Healpix4'' and
``Healpix8'' refer to iterative analyses with 4 and 8 steps, respectively.}
\label{error_rms}
\end{figure}

\begin{figure}
\begin{center}
\includegraphics[width=0.9\columnwidth]{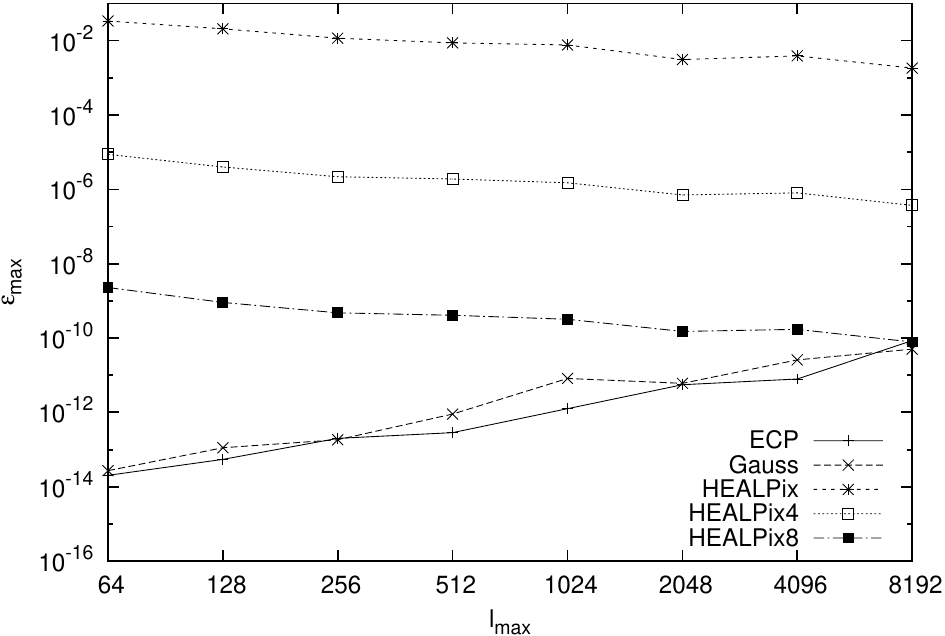}
\end{center}
\caption{$\varepsilon_{\text{max}}$ for \libpsht\ transform pairs on ECP, Gaussian
and HEALPix grids for various \lmax. Every data point is the maximum error value
encountered in transform pairs of all supported spins. ``Healpix4'' and
``Healpix8'' refer to iterative analyses with 4 and 8 steps, respectively.}
\label{error_max}
\end{figure}

A variety of such tests was performed for different grid geometries, \lmax,
and spins; the results are shown in Figs.\ \ref{error_rms} and \ref{error_max}.


For ECP and ``Gaussian'' grids, each ring consisted of $2\lmax+1$ pixels, and the
grids contained $2\lmax+2$ and $\lmax+1$ rings, respectively. The $\vec\vartheta_y$
for the Gaussian grid were chosen to coincide with the roots of the Legendre
polynomial $P_{\lmax+1}(\vartheta)$. Under these conditions, and with appropriately
chosen pixel weights (taken from \citealt{driscoll-healy-1994} and
\citealt{doroshkevich-etal-2005}), the transform pair should reproduce the original \alm\ exactly, were
it not for the limited precision of floating-point arithmetics. In fact, the
errors measured for SHTs on those two grids are extremely small and close
to the theoretical limit.

For the transform pairs on HEALPix grids, a resolution parameter of
$\nside=\lmax/2$ was chosen. Here, the recovered ${}_s\hat a_{lm}$
will only be an approximation of the original ${}_sa_{lm}$; if higher accuracy
is required, this approximation can be improved by Jacobi iteration, as described
in the HEALPix documentation.
The figures show clearly that $\varepsilon_{\text{rms}}$ is around $10^{-3}$
for the simple transform pair and can be reduced dramatically by iterated
analysis.



\subsection{Time consumption}

\subsubsection{Relative cost of SHTs}
\begin{table}
\caption{Single-core CPU time for various SHTs with $\lmax=2048$.}
\begin{center}
\begin{tabular}{lcccc}
\hline\hline
Direction\vphantom{\Large I}&Spin   & ECP & Gaussian & HEALPix\\
\hline
forward  &0     & 15.14\,s  & \phantom{0}7.89\,s  & 13.11\,s\\
backward &0     & 14.74\,s  & \phantom{0}7.53\,s  & 12.65\,s\\
forward  &1     & 29.49\,s  & 15.36\,s & 25.05\,s\\
backward &1     & 27.83\,s  & 14.30\,s & 23.85\,s\\
forward  &2     & 31.91\,s  & 16.91\,s & 28.15\,s\\
backward &2     & 30.33\,s  & 15.60\,s & 26.51\,s\\
\hline
\end{tabular}
\end{center}
\tablefoot{The ECP grid had $2\lmax+1$ pixels per ring and $2\lmax+2$ rings, the Gaussian grid
had $2\lmax+1$ pixels per ring and $\lmax+1$ rings, and the \nside\ parameter
of the HEALPix grid was $\lmax/2$.}
\label{perfnumbers}
\end{table}
Table \ref{perfnumbers} shows the single-core CPU times for SHTs of different
directions, spins and grids, but with identical band limit, in order to
illustrate their relative cost. It is fairly evident that forward and backward
SHTs with otherwise identical parameters require almost identical time, with
the $\alm\rightarrow\text{map}$ direction being slightly slower. Since this
relation also holds for other band limits, only the timings for one direction
will be shown in the following figures, in order to reduce clutter.

SHTs with $|s|>0$ take roughly twice the time of corresponding scalar ones;
here the $s=2$ case is slightly more expensive than $s=1$, because computing
the ${}_2F^\pm_{lm}$ from $\lambda_{lm}$ requires more operations than computing
the ${}_1F^\pm_{lm}$.

The timings also indicate the scaling of the SHT cost with \ntheta, which is
practically equal for HEALPix and ECP grids, but half as high for the Gaussian
grid.

These timings, as well as all other timing results for \libpsht\ in this paper,
were obtained using SSE2-accelerated code. Switching off this feature will
result in significantly longer execution times; on the benchmark computer
the factor lies in the range of 1.6 to 1.8.

\subsubsection{Scaling with \lmax}
\begin{figure}
\begin{center}
\includegraphics[width=0.9\columnwidth]{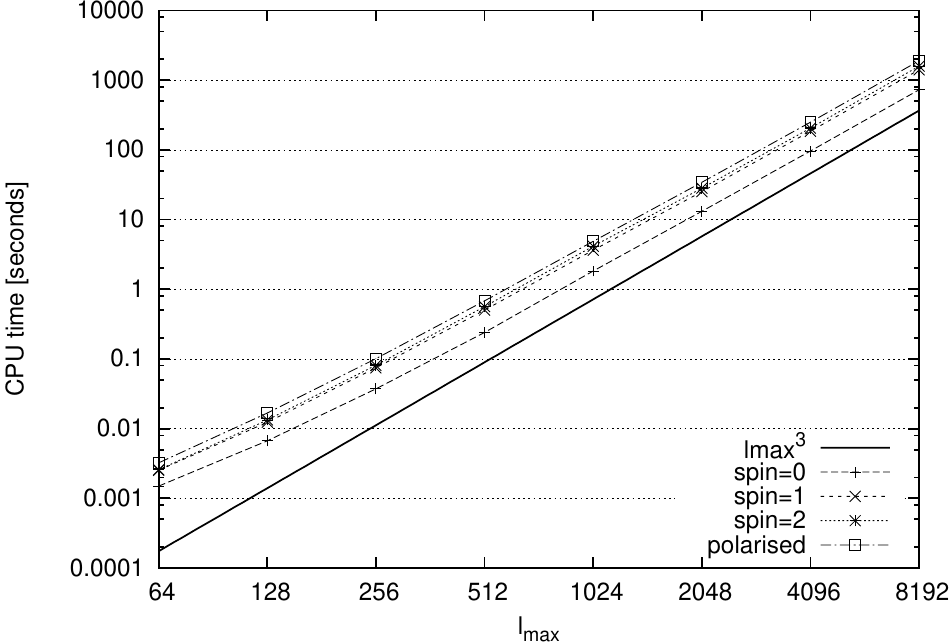}
\end{center}
\caption{Timings of single-core forward SHTs on a HEALPix grid with
$\nside=\lmax/2$. The annotation ``polarised'' refers to a combined SHT of
spin 0 and 2 quantities, which is a very common case in CMB physics.}
\label{l_scaling}
\end{figure}
\begin{figure}
\begin{center}
\includegraphics[width=0.9\columnwidth]{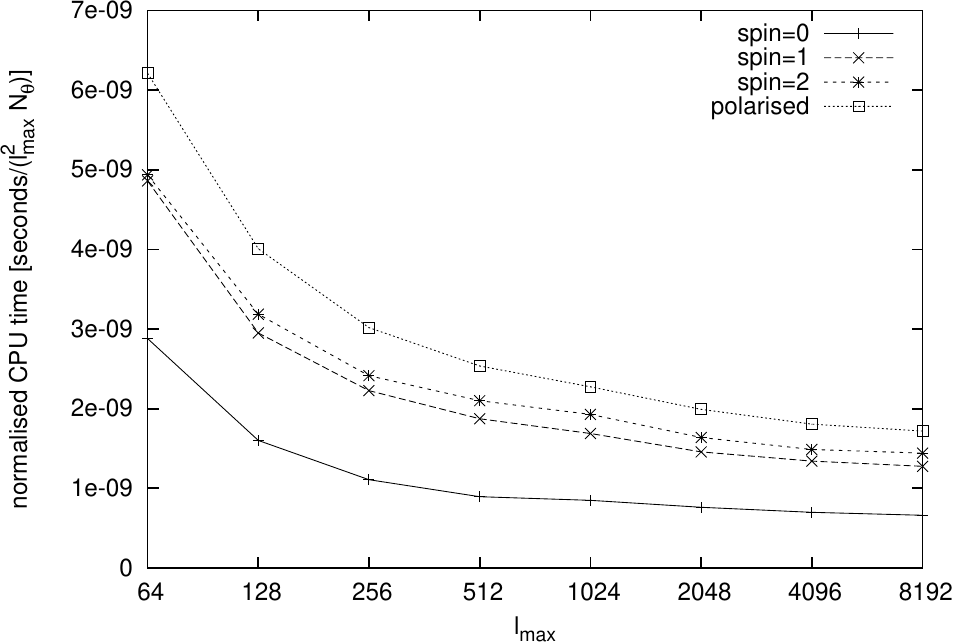}
\end{center}
\caption{Same as Fig.~\ref{l_scaling}, but with CPU time divided by $\lmax^2\ntheta$}
\label{norm_l_scaling}
\end{figure}

SHTs of different spins and a wide range of band limits were performed on a
single CPU core, in order to assess the correlation between CPU time and \lmax;
the results are shown in Fig.~\ref{l_scaling}. As expected, the time consumption
scales almost with $\lmax^3$; deviations from this behaviour only become apparent
for lower band limits, where SHT components of lower overall complexity
(like the FFTs, computation of the $A_{lm}$ and $B_{lm}$ coefficients and
initialisation of data structures) are no longer negligible.
Switching to a normalised representation of the CPU time, as was done in
Fig.~\ref{norm_l_scaling}, makes this effect much more evident; even for transforms
of very high band limit the curves are not entirely horizontal, which would indicate
a pure \compl{\lmax^2\ntheta} algorithm.

\subsubsection{Performance comparison with Fortran HEALPix}
\begin{figure}
\begin{center}
\includegraphics[width=0.9\columnwidth]{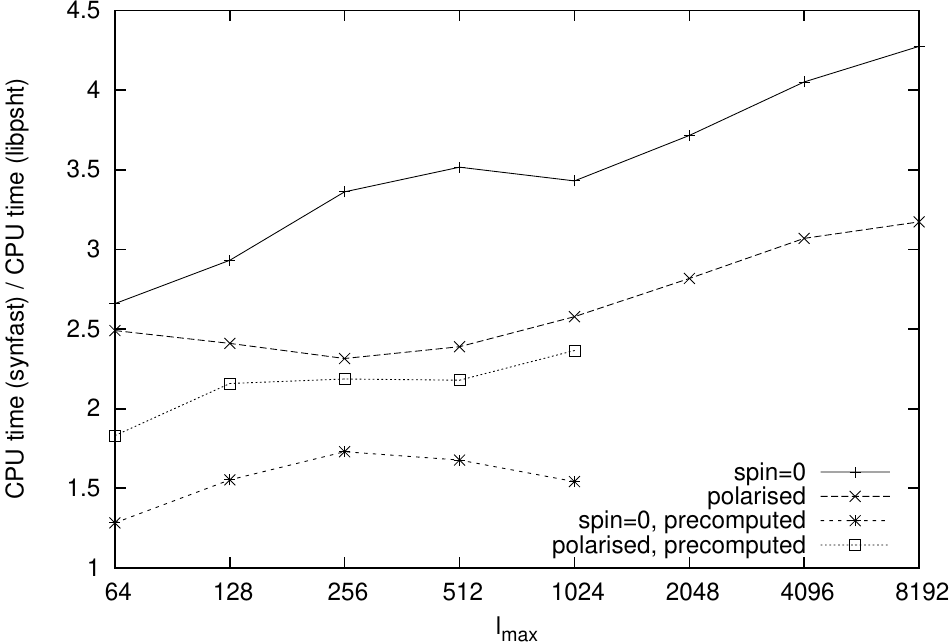}
\end{center}
\caption{CPU time ratios between the Fortran HEALPix \texttt{synfast} code and
\libpsht\ for various backward SHTs on a HEALPix grid with $\nside=\lmax/2$.}
\label{hpcomp}
\end{figure}
Fig.\ \ref{hpcomp} shows the ratio of CPU times for equivalent SHTs carried
out using the \texttt{synfast} facility of Fortran HEALPix and \libpsht, using
identical hardware resources. It allows several deductions:
\begin{itemize}
\item In all tested scenarios, \libpsht's implementation is significantly faster,
  even when precomputed coefficients are used with \texttt{synfast}.
\item The performance gap is markedly wider for $s=0$ transforms, which
  indicates that \libpsht's $l$-recursion is the part which has been accelerated
  by the largest factor in comparison to the Fortran HEALPix SHT library.
\item The polarised \texttt{synfast} SHTs only show a marginal speed improvement
  when using precomputed coefficients; for the scalar transforms, the difference
  is much more noticeable. Although this observation has no direct connection
  to \libpsht, it is nevertheless of interest: most likely the missing acceleration
  is caused by saturation of the memory bandwidth, as discussed in Sect.\ \ref{precompute},
  and confirms the decision not to introduce such a feature into \libpsht.
\end{itemize}

\subsubsection{Scaling with number of cores}
\label{openmp_scaling}
\begin{figure}
\begin{center}
\includegraphics[width=0.9\columnwidth]{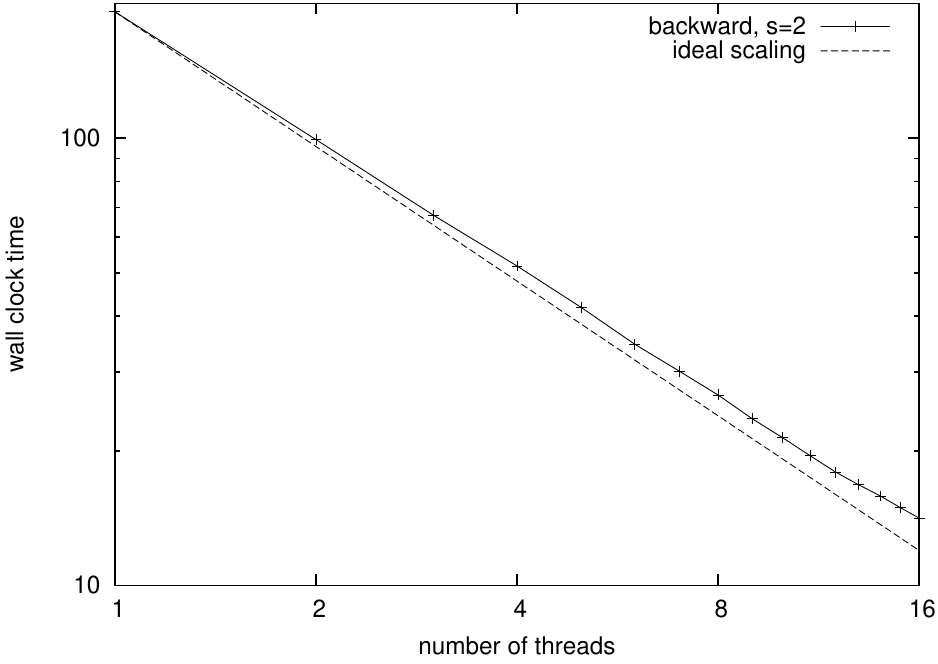}
\end{center}
\caption{Elapsed wall clock time for a backward SHT with $s=2$, $\lmax=4096$,
and $\nside=2048$ on a HEALPix grid, run with a varying number of OpenMP threads.}
\label{ompscale}
\end{figure}
As illustrated in Fig.~\ref{loopstruct}, all CPU-intensive parts of \libpsht's
transform algorithm are instrumented for parallel execution on multicore
computers using OpenMP directives. Fig.~\ref{ompscale} shows the
wall clock times measured for a single SHT that was run using a varying number
of OpenMP threads. The maximum number of threads was limited to 16 (in contrast
to the maximum useful value of 24 on the benchmark computer), because the machine
was not completely idle during the \libpsht\ performance measurements, and
the scaling behaviour at larger numbers of threads would have been influenced
by other running tasks.

It is evident that the overhead due to workload distribution among several
threads is quite small: for the 16-processor run, the accumulated wall clock time
is only 18\% higher than for the single-processor SHT. An analogous experiment
was performed with the Fortran HEALPix code \texttt{synfast}; here the overhead
for 16 threads was around 35\%.

\subsubsection{Comparison with theoretical CPU performance}
For this experiment, the library's code was enhanced to count the number of
floating-point operations that were carried out during the SHT. This instrumentation
was only done in the parts of the code with \compl{\lmax^2N_\vartheta} complexity,
i.e.\ everything related to creation and usage of the \lamlm. Operations that
were necessary only for the purpose of numerical stability (e.g.\ due to the
range-extended floating-point format) were not taken into account.
Dividing the resulting number of operations by the CPU time for a single-core
uninstrumented run of the same SHT provides a conservative estimate for
the overall floating-point performance of the algorithm.
Measuring $s=0$ and $s=2$ $\alm\rightarrow\text{map}$ SHTs on a HEALPix grid
with $\nside=1024$ and $\lmax=2048$ resulted in 2.1GFlops/s and 4.1GFlops/s,
respectively. This corresponds to 0.88 and 1.71 floating-point operations per
clock cycle, or 22\% and 43\% of the theoretical peak performance.
Both of these percentages are quite
high for an implementation of a nontrivial algorithm, which can be seen as a hint
that there is not much more room for further optimisation on this hardware architecture.
In other words, further speedups can only be achieved by a fundamental change
in the underlying algorithm.

There are several reasons for the obvious performance discrepancy between scalar
and $s=2$ transforms. To a large part it is probably
caused by the internal dependency chains of the $\lambda_{lm}$ recursion: even though
the CPU could in principle start new operations every clock cycle, it has to
wait for some preceding operations to finish first (which takes several cycles),
because their results are needed in the next steps.
When computing a recurrence, the CPU is therefore spending a significant
fraction of time in a waiting state (see, e.g., \citealt{fog-2010}).
For the $s=2$ transform this is much less of an issue, since many more
floating-point operations must be carried out, which are not interdependent,
and the time spent on the $l$-recursion becomes subdominant.

Other reasons for the lower performance of the recursion are the necessity to
deal with extended floating-point numbers and its rather high ratio of memory accesses
compared to arithmetic operations.

\subsubsection{Gains by using simultaneous transforms}
The CPU time savings achievable by performing multiple SHTs together instead of one after
another were measured on a HEALPix grid with $\nside=1024$ and $\lmax=2048$.
Table \ref{simul_speedup} summarises the measured CPU times for
an unsystematic selection of simultaneous and successive transforms, as well
as the speedup.

\begin{table}
\caption{Speedup for simultaneous vs.\ consecutive SHTs on a HEALPix grid
with $\nside=1024$ and $\lmax=2048$.}
\begin{tabular}{lrrr}
\hline\hline
\vphantom{\Large I}Transforms & $T_{\text{simul}}$/s & $T_{\text{separate}}$/s & Speedup\\
\hline
1a0                     & 12.65 &  12.65 & 1.00\\
2a0                     & 18.06 &  25.30 & 1.40\\
3a0                     & 23.47 &  37.95 & 1.62\\
5a0                     & 34.77 &  63.25 & 1.82\\
10a0                    & 62.56 & 126.50 & 2.02\\
3a0+3m0                 & 41.72 &  77.46 & 1.85\\
1a0+1a1+1a2+1m0+1m1+1m2 & 78.79 & 129.44 & 1.64\\
4m2                     & 63.68 & 112.40 & 1.77\\
\hline
\end{tabular}
\tablefoot{The transforms are
specified in an abbreviated fashion, where ``5a2'' signifies
``5 $\alm\rightarrow\text{map}$ SHTs of spin 2'', and ``3m0'' stands for
``3 $\text{map}\rightarrow\alm$ SHTs of spin 0''. The runs were performed on a
single CPU core.}
\label{simul_speedup}
\end{table}

The results show that for a large number of simultaneous SHTs the computation
time can be cut roughly in half; this is the same asymptotic limit that also
holds for SHTs with completely precomputed \lamlm\ (see Sect.\ \ref{precompute}),
but the presented approach has dramatically lower memory needs and can therefore
be used for higher-resolution transforms. It is also evident that forward and backward
transforms, as well as transforms of different spins, can be freely mixed while
still preserving the performance increase.

\subsection{Memory consumption}
\begin{figure}
\begin{center}
\includegraphics[width=0.9\columnwidth]{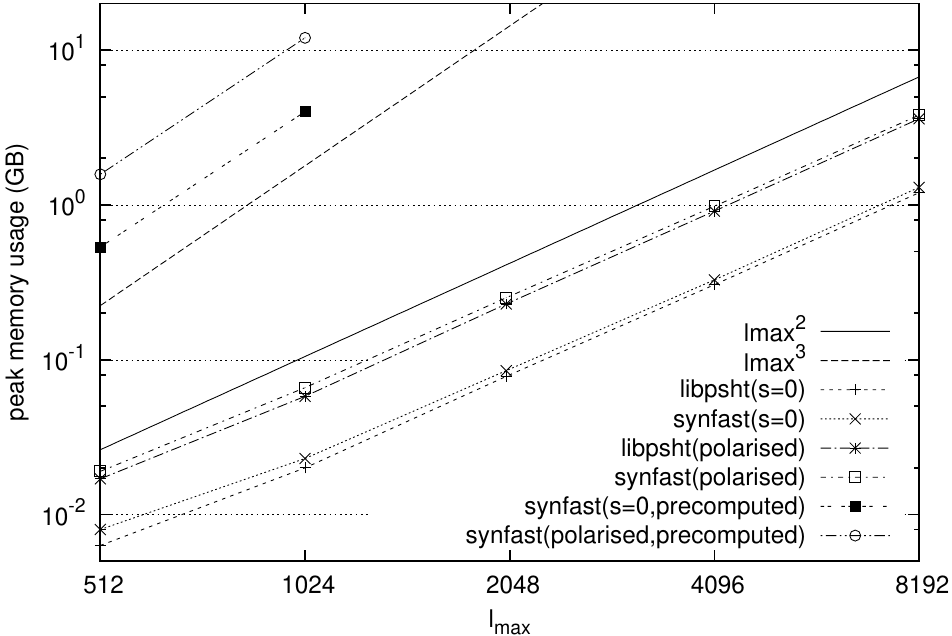}
\end{center}
\caption{Memory consumption of various SHTs performed with \libpsht\ and
\texttt{synfast} on a HEALPix grid with $\nside=\lmax/2$. Input and output data
are stored in single-precision format.}
\label{memsize}
\end{figure}
Fig.~\ref{memsize} shows the maximum amount of memory allocated during a variety
of SHTs carried out using \libpsht\ and the Fortran HEALPix \texttt{synfast}
facility. Since the memory required for forward and backward transforms is
essentially equal, only one direction has been plotted. The sizes of SHTs with
$\lmax<512$ could not be measured reliably due to their short running time and
are therefore not shown.

As can be seen, \libpsht\ needs slightly less memory than \texttt{synfast} for equivalent
operations; overall, the scaling is very close to the expected $\lmax^2$ if
no precomputed \lamlm\ are used. For comparison purposes, a few data points
for \texttt{synfast} runs using precomputed scalar and tensor \lamlm\ were also
plotted; they clearly indicate the extremely high memory usage of these
transforms, which scales with $\lmax^3$.

During the tests with multiple simultaneous threads it was observed that
increasing the number of cores does not have a significant influence on the total
required memory; for the concrete test discussed in Sect.\ \ref{openmp_scaling},
memory usage for the run with 16 threads was only 2\% larger than for the
single-threaded run.


\section{Conclusions}
Looking back at the goals outlined in Sect.\ \ref{goals} and the results
of the benchmark computations above, it can be concluded that the current version
of the \libpsht\ package meets almost all specified requirements. It implements
SHTs that have the same or a higher degree of accuracy as the other available
implementations, can work on all spherical grids relevant to CMB studies, and
is written in standard C, which is very widely supported.
The employed algorithms are significantly more efficient than those published
and used in this field of research before; their memory usage is economic and
allows transforms whose input and output data occupy almost all available space.
When appropriate, vector capabilities of the CPUs, as well as shared memory
parallelism are exploited, resulting in further performance gains.

Despite this range of capabilities, the package only consists of approximately
7000 lines of code, including the FFT implementation and in-line documentation
for developers; except for a C compiler, it has no external dependencies.

\Libpsht\ has been integrated into the simulation package of the \textit{Planck} mission
(\textit{Level-S}, \citealt{reinecke-etal-2006}), where it is interfaced with a development version of the C++ HEALPix
package\footnote{The next major release of HEALPix C++ will use \libpsht.}
and a Fortran90 code performing SHTs on detector beam patterns; this
demonstrates the feasibility of interfacing the library with C++ and Fortran
code.

There are two areas in which \libpsht's functionality should probably be extended:
it currently  does not provide transforms for spins larger
than 2, and there is no active support for distributing SHTs over several
independent computing nodes. As was mentioned in Sect.\ \ref{tensorSHT},
addressing the first point is probably not too difficult, and enhancing
the library with ${}_sY_{lm}$ generators based on Wigner $d$ matrix elements
is planned for one of the next releases. Regarding the second point, a combination
of \libpsht's central SHT functionality with the parallelisation strategy
implemented by Radek Stompor's \texttt{s2hat} library appears very promising and
is currently being investigated.

\Libpsht\ is distributed under the terms of the GNU General Public License
(GPL) version 2 (or later versions at the user's choice); the most recent
version, alongside online technical documentation can be obtained at the URL
\url{http://sourceforge.net/projects/libpsht/}.

\begin{acknowledgements}
Some of the results in this paper have been derived using the HEALPix
\citep{gorski-etal-2005}
and GLESP \citep{doroshkevich-etal-2005} packages.
MR is supported by the German Aeronautics Center and Space Agency (DLR), under
program 50-OP-0901, funded by the Federal Ministry of Economics and
Technology.
I am grateful to Volker Springel, Torsten Ensslin, J\"org Rachen and
Georg Robbers for fruitful discussions and valuable feedback on drafts of
this paper.
\end{acknowledgements}

\bibliographystyle{aa}
\bibliography{planck}
\end{document}